\documentclass[11pt]{article}
\usepackage[T2A]{fontenc}
\usepackage{booktabs}
\usepackage[utf8]{inputenc}
\usepackage{siunitx}
\usepackage{xcolor}
\usepackage{hyphenat}
\usepackage[paperwidth=170mm,paperheight=240mm,centering,height=186mm,width=120mm]{geometry}
\usepackage{geometry}
\usepackage{graphicx}
\usepackage{amsmath}
\usepackage{amssymb}
\begin{document}

\title{On the Possibility of Direct Detection of the Emission of Microlens MOA-2011-BLG-191/OGLE-2011-BLG-0462 - a Probable Black Hole
}
\author{L.Chmyreva$^1$, G.Beskin$^{1,2}$}
\date{%
    $^1$Special astrophysical observatory of RAS,
  Nizhny Arkhyz, Russia\\%
    $^2$Kazan (Volga Region) Federal University, Kazan, 420008 Russia\\[2ex]%
    %\today
}

\maketitle

\begin{abstract}
We discuss the observational manifestations of an isolated stellar mass black hole - the recently discovered microlens MOA-2011-BLG-191/OGLE-2011-BLG-0462. The data available for this object are used to calculate the density, temperature, and sound speed in its local interstellar medium, as well as estimate its velocity. We obtain the accretion rate and luminosity of the object, and construct its theoretical spectrum. A comparison of the spectrum with the sensitivity levels of current and future instruments in different frequency ranges has shown that direct detection of the emission from this black hole is possible for several future observing missions.
\end{abstract}
 
\section{Introduction}

Stellar mass black hole (BH) candidates are formed in the end of evolution of stars with masses \mbox{$M \gtrsim 25 M_\odot$}. According to current evolutionary predictions, the number of isolated BHs in our Galaxy amounts to approximately $10^8$ \cite{2019ApJ...885....1W}. However, detecting isolated black holes is very difficult in comparison with those that are members of x-ray binaries and whose masses can be estimated. At the same time, data from regions in direct proximity of the event horizon, which is a generic attribute of a BH \cite{2008AdSpR..42..523B}, can probably be obtained only for an isolated BH: in this case the accretion rate is low and the event horizon is not obscured by the accreting matter. Interstellar gas is captured by the gravitational field of a BH of any type at the radius
$R_c=\frac{2GM}{V^2+c_s^2},$ which determines the accretion rate
\cite{1944MNRAS.104..273B, 2005A&A...440..223B,
1992ans..book.....L}. For the expected BH velocity
(< $100$~km\,s$^{-1}$) and a typical mass of \mbox{$5$--$10$
$M_\odot$}, $R_c$ exceeds the size of the ergosphere (event horizon) by a factor of millions, and the gravitational field can be described in terms of Newton's approximation. At the same time, the nature of the flares generated by beams of electrons in the direct vicinity of a BH will depend drastically on its type. Manifestations of nontrivial structure, statistics, polarization features of such events can be expected for a Kerr black hole with axial symmetry and an ergosphere \cite{1981cfgr.book.....K,
1986UsFiN.148..393D}. Detecting and studying such events with maximal temporal resolution would allow us to obtain information on the structure of space-time near a black hole, as well as prove the existence of an event horizon and ergosphere, i.e. generic features of a BH.

Accretion onto an isolated BH, regardless of its metrics, is usually spherical; the spectrum for this accretion type was first derived by Shvartsman \cite{1971SvA....15..377S}. Luminosity for this type of accretion remains almost constant in a wide range of frequencies 
\mbox{($10^{14}$--$10^{20}$ Hz)}, and the spectrum lacks lines
\cite{1971SvA....15..377S, 1974Ap&SS..28...45B,
1975A&A....44...59M, 1982ApJ...255..654I}. For typical interstellar medium parameters and expected BH velocities and masses, the latter will manifest themselves as objects with luminosities of
\mbox{$10^{28}$--$10^{34}$ erg\,s$^{-1}$} and continuous spectra. Their common property would be variable emission with an amplitude ranging from fractions of a percent to ten percent; the duration of individual flares would be~$10^{-6}$ to~$10^{-3}$~s. At the same time these objects may be variable on time scales of months and years, which is due to the inhomogeneity of the interstellar medium through which a BH moves (see
\cite{2008AdSpR..42..523B, 2005A&A...440..223B} and references therein). These ideas were the foundation of the ``MANIA'' experiment dedicated to the search for isolated stellar mass BHs throughout the entire northern sky, based on the detection of fast variations of the mentioned emission in the optical range \cite{1971SvA....15..377S,
2008AdSpR..42..523B, 2005A&A...440..223B,1977SoSAO..19....5S,
1989Afz....31..457S, 1989SvAL...15..145S, 2010AJ....139..390P}.

We attempted to narrow down the search area and carry out an analysis in regions with an inherently high probability of locating a BH in our previous papers,
\cite{2010AstL...36..116C, 2022AstBu..77...65C}, where we searched for candidate objects that were companions in disrupted binaries.  

An important method of searching for isolated BHs consists of studying their x-ray and radio emission. The contribution of isolated BHs to the detected x-ray radiation can be comparable to that of neutron stars. Additionally, the hard spectral component is highly variable due to the emission of blobs of non-thermal electrons \cite{2005A&A...440..223B}. Estimates show that the sensitivity of modern x-ray observatories (NuStar, Spektr-RG) is sufficient for detecting BHs
(see, e.g., \cite{2002MNRAS.334..553A,
2018MNRAS.477..791T}). Isolated BHs may also be the sources of gamma and radio emission in molecular clouds and regions of cold neutral hydrogen \cite{2012MNRAS.427..589B,
2005MNRAS.360L..30M}. Gravitational wave observations make it possible to detect black hole and/or neutron star mergers and determine the mass of the formed BH, but the low position accuracy (degrees) does not allow one to investigate the event horizon directly (see, e.g.,
\cite{2016PhRvX...6d1015A, 2016ApJ...818L..22A}). The possibility of BH detection in the radio within the framework of some planned future missions was studied in detail in \cite{2019MNRAS.488.2099T}. Model computations of \cite{2013MNRAS.430.1538F, 2021MNRAS.505.4036S} have shown that searching for BHs in the radio is even more feasible considering the sensitivity of such projects as SKA. The prospects of searching for isolated BHs in the central molecular zone of the Galaxy in the sub-millimeter and infrared spectral regions by means of space observatories are discussed in
\cite{2019MNRAS.489.2038I}.

However, the only method that allows one to obtain a direct mass estimate for an isolated BH candidate (we remind the reader that the only direct evidence of an Einsteinian black hole detection is the presence of an event horizon in the massive compact candidate object) is astrometric microlensing. Modeling has shown \cite{2016ApJ...830...41L} that this method is achievable in observations, the results of which, in particular, are presented in \cite{ 2022arXiv220201903L,
2022arXiv220113296S}. And while the compact lens object in the first study may be either a black hole and a neutron star, the mass estimate \mbox{($7.1\pm1.3~M_{\odot}$)} and luminosity limits for the lens in 
\cite{2022arXiv220113296S} leave no doubt that the object, \mbox{MOA-2011-BLG-191/OGLE-2011-BLG-0462}, is probably a black hole. However, as we emphasized repeatedly \cite{2005A&A...440..223B}, only the detection of observational manifestations of an event horizon would become the final argument for identifying the object in question as a black hole. In any case,
\mbox{MOA-2011-BLG-191/OGLE-2011-BLG-0462} is the best candidate for searching for and studying an event horizon. To that end, in this work we evaluate the capabilities of existing and future astronomical instruments for solving this task. Based on the mass and velocity estimates for the BH and its local interstellar medium parameters, we obtain, within the low rate spherical accretion model \cite{1944MNRAS.104..273B, 1952MNRAS.112..195B,
1971SvA....15..377S}, the spectral energy distribution for the emission of matter accreting onto the lens. Additionally, its stationary (total) and flare components are separated and compared to the sensitivity (existing and planned) of telescopes in different wavelength regions. Our analysis shows that one cannot rule out a possible detection of observational manifestations of the event horizon of
\mbox{MOA-2011-BLG-191/OGLE-2011-BLG-0462}.

\section{Observational Manifestations of Gravitational Lens
 MOA-11-191/OGLE-11-0462, a Probable Black Hole}

Sahu et al. \cite{2022arXiv220113296S} use the Hubble Space Telescope (HST) to carry out precise astrometric observations of the background source star over the course of a long-term microlensing event
MOA-2011-BLG-191/OGLE-2011-BLG-0462
(\mbox{MOA-11-191/OGLE-11-0462}). Frames taken with HST over a six-year interval for eight epochs of observations show a clear relativistic astrometric shift of the visible position of the lensed background star. Based on this result, the star's light curve and its parallax estimate, the mass of the lens was determined as
\mbox{$7.1\pm1.3~M_{\odot}$} \cite{2022arXiv220113296S}, which exceeds significantly the maximum masses of white dwarfs and neutron stars. The distance was estimated as \mbox{$1.58\pm0.18$~kpc}, and the transverse velocity is equal to about \mbox{$45\pm
5$~km\,s$^{-1}$}. In the next section, these estimates are used to determine the luminosity and spectrum of the gravitational lens in an assumption that it is an isolated black hole.

\subsection{Characteristics of the probable black hole MOA-11-191/OGLE-11-0462 and its surroundings}

Our previous papers, \cite{2008AdSpR..42..523B,
2005A&A...440..223B}, are dedicated to a theoretical analysis of observational manifestations of isolated stellar mass black holes within the framework of a spherical interstellar gas accretion model. We used the results of that analysis to search for black hole candidates~--- the former companions of radio pulsars in disrupted binaries \cite{2022AstBu..77...65C}.

According to spherical accretion models \cite{1952MNRAS.112..195B,
1971SvA....15..377S}, the luminosities $L$ of halos around isolated BHs are determined by their masses and velocities, as well as the temperature and density of the interstellar medium in their vicinities, specifically
\cite{1944MNRAS.104..273B,1971SvA....15..377S,2005A&A...440..223B}:

\begin{equation}
L=9.6\times10^{33}M^{3}_{10}n^2(V^2+c^2_s)^{-3}_{16}\;\;\text{erg s$^{-1}$},
\label{L}
\end{equation}
where $M_{10}$ is the BH mass normalized to \mbox{$10 M_\odot$},
$n$ is the medium density in cm$^{-3}$, and $V$ and $c_s$ are the total BH velocity and sound speed normalized to $16$~km\,s$^{-1}$. Note that the numerical coefficient here is several times higher than the coefficients in similar expressions from
\cite{1982ApJ...255..654I} and \cite{1974Ap&SS..28...45B}. This is due to the use in \cite{2005A&A...440..223B} of a more detailed electron heating model, taking into account the influence of a magnetic field on this process.

In order to determine the interstellar medium density in the vicinity of
\mbox{MOA-11-191/OGLE-11-0462} we used the empirical $E_{g-r}(\mu)$ dependences derived from the data of the Galactic 3D dust distribution map
\cite{2019ApJ...887...93G}. Here $E_{g-r}$ is the interstellar reddening and $\mu$ is the distance modulus. The density $n$ is the derivative of the hydrogen column density $N_{\rm H}$ with respect to distance~$D$. Using the dependences  \linebreak $N_{\rm
H}=6.86\times 10^{21} E_{B-V}$~~\cite{2009MNRAS.400.2050G}~ and
$E_{B-V}=0.884 E_{g-r}$ \cite{2019ApJ...887...93G}, we derive
\mbox{$N_{\rm H}=f(\mu)=6.06\times 10^{21} E_{g-r}(\mu)$}.
Differentiating this function and taking into account the relation
$\mu=5\lg(\frac{D}{10})$, we obtain after simple transformations an expression for the local medium density:
\begin{equation}
n=\dfrac{dN_{\rm
H}}{dD}=1.3\times10^{21}\times10^{-\mu/5}\dfrac{dE_{g-r}(\mu)}{d\mu}.
\label{n}
\end{equation}

\begin{figure}[ht!]
   \centering
    \includegraphics[width=1.0\textwidth]{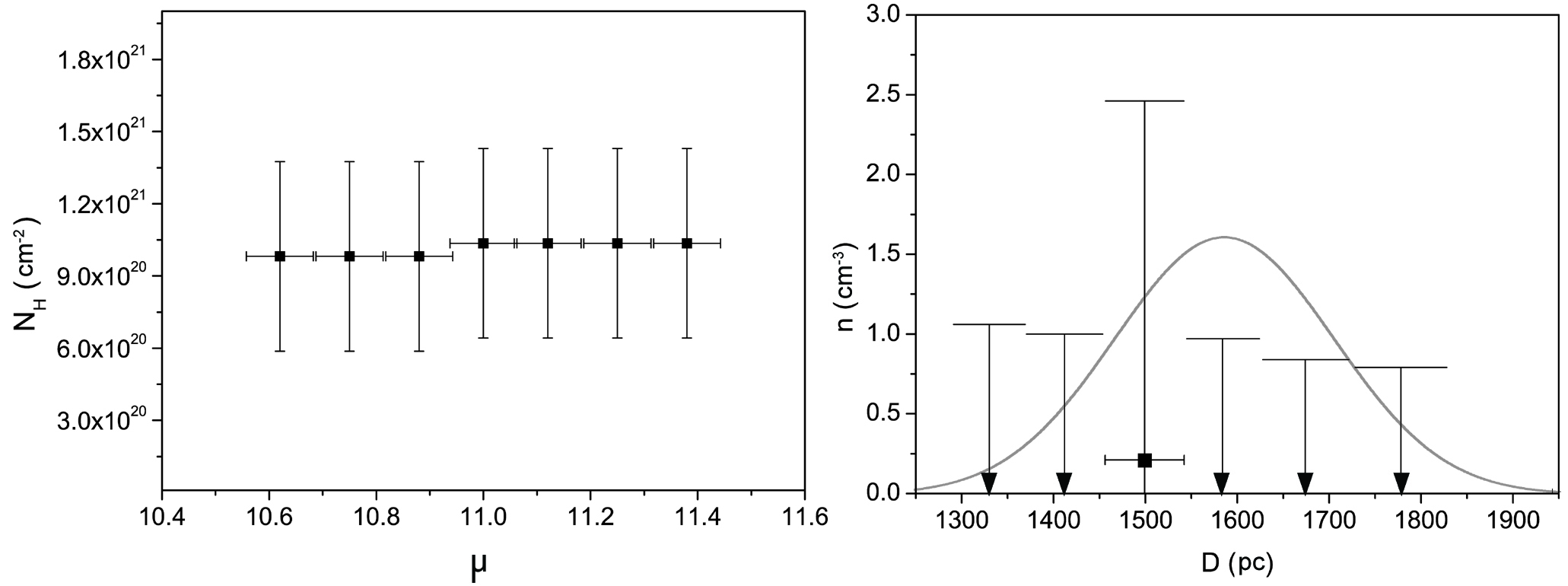}
\caption{\small{Left: column density $N_{\rm H}$ as a function of distance modulus $\mu$ for the probable black hole
\mbox{MOA-11-191/OGLE-11-0462}. The uncertainties correspond to the
$1\sigma$ level according to the distribution from \cite{2019ApJ...887...93G}.
Right: interstellar medium density $n$ as a function of distance in the direction of the lens for the lens distance estimate of
\mbox{$1.58\pm0.18$}~kpc (the curve shows the normal distribution of that quantity). The arrows show the upper limits corresponding to a $99\%$ confidence level.}}
\label{fig:n}
\end{figure}

Fig.~\ref{fig:n} shows the dependence of the interstellar medium density from expression (\ref{n}) near MOA-11-191/OGLE-11-0462 on distance. The weighted average \mbox{$\overline{n}=0.7\pm0.4$} for the range of acceptable distances \mbox{$1.58\pm0.18$~kpc} is used as its local estimate. In particular, $\overline{n}\in(\overline{n}_{min},\overline{n}_{max})$, where $\overline{n}_{min}=\frac{\sum w n}{\sum w}$ and $\overline{n}_{max}=\frac{\sum w n+\sum w n_{0}}{\sum w}$.
Here $n$ represents the individual density values, $w$ is their probability from the normal distribution in Fig.~\ref{fig:n}, and
$n_0$ is the upper density limit estimate.

The sound speed in the BH vicinity was determined using the standard formula:  $c_s=\sqrt{\frac{\gamma kT}{m_p}}$ where $\gamma$ is the hydrogen adiabatic index, $m_p$ is the proton mass, and the temperature $T$ was determined from the empirical dependence $T(n)$ \cite{1981sss..book..265B}. Here the range of its values was determined by the interval of acceptable densities $\overline n$. 

The transverse velocity $V_{\rm tr}$, obtained from the observed proper motion \cite{2022arXiv220113296S}, is an estimate of the minimum full spatial velocity of the BH and places a lower limit on the range of acceptable $V$ values, i.e. $V>V_{\rm tr}$.

\subsection{Luminosity and spectrum of the probable black hole~---
gravitational lens MOA-11-191/OGLE-11-0462}

The magnitudes of the lensed source outside of the brightness increase interval are $m_V=21.946\pm0.014$  and
$m_I=19.581\pm0.012$ in the $V$ and $I$ bands, correspondingly \cite{2022arXiv220113296S}. No light from the lens was detected even in the data from the last epoch of observations ($6.1$~years after the peak), when the angular separation between the lens and the source was approximately $42.6$~mas. Based on this, the authors
\cite{2022arXiv220113296S} estimate the upper limit for the luminosity of the lens at a level of $1\%$ of the source, i.e. the \mbox{$V$-band} magnitude of the lens should be fainter than
$m_V+5^m\simeq27^m$ (for an exposure of one hour, the HST can register sources up to $V=27.9^m$ \cite{2022wfci.book...14D}, and thus its sensitivity may be sufficient for detecting the object once it has moved away from the source).

In accordance with the computations described above, the interstellar medium density estimate in the vicinity of the BH lies in the interval from $0.06$ to $1.3$~cm$^{-3}$, and that for the temperature~--- from $200$~K to $12\,500$~K, which gives the speed of sound in the range of approximately $1.5$ to $12$~km\,s$^{-1}$. The extinction $A_{V}\sim3.1 E_{B-V}$ for the object, calculated from the data of the dust distribution map \cite{2019ApJ...887...93G} with account for the above mentioned transformations, amounts to
$0.62^{+0.32}_{-0.06}$, where the errors correspond to the uncertainties of distance ($1.58\pm0.18$~kpc).

\begin{figure}[ht!]
   \centering
    \includegraphics[width=0.7\textwidth]{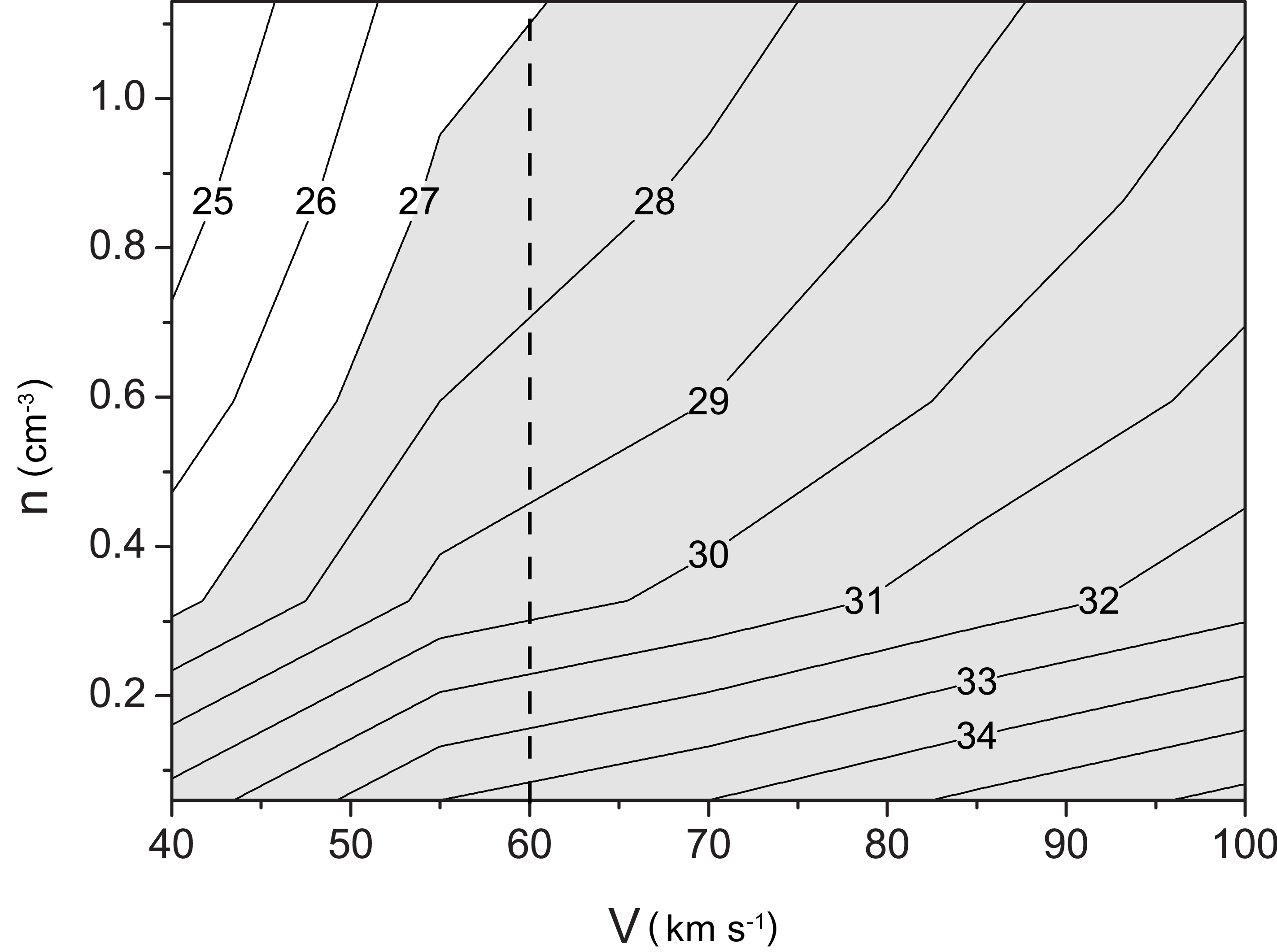}
\caption{\small{The $m_{V}$ magnitude estimates for
\mbox{MOA-11-191/OGLE-11-0462} (stripes with the corresponding values) for interstellar medium density
$n\in(0.06,1.3)$ cm$^{-3}$ and total velocity
$V\in(40,100)$ km s$^{-1}$. The grey region shows magnitudes fainter than $27^{\rm m}$, which corresponds to the approximate lower limit estimate for $m_{V}$ \cite{2022arXiv220113296S}. For a velocity of
\mbox{$V \gtrsim60$~km\,s$^{-1}$} (vertical line) this estimate is valid for almost the entire range of $n$.}}
\label{fig:vn}
\end{figure}

Velocity $V\in(40,100)$ and density $n\in(0.06,1.3)$ are the parameters that contribute the most to the uncertainty of $m_{V}$. The scatter of apparent magnitudes when varying these parameters within the given intervals is about $6^m$ and $6.5^m$
correspondingly (Fig.~\ref{fig:vn}). The BH mass has a smaller influence: the scatter of $m_{V}$ amounts to about $1.2^m$ for a mass interval of $5.8$ to $8.4\,M_\odot$. The distance and temperature uncertainties have an even smaller effect: the approximate scatter for them is $\sim0.5^m$ and $\sim0.3^m$.

As is evident from Fig.~\ref{fig:vn}, the object would be at its brightest in the low velocity and high surrounding medium density scenario: the magnitude estimate in this case exceeds the acceptable limit derived from the data of \cite{2022arXiv220113296S}. Our modeling has shown that the limit is valid for a lower density and/or higher velocity. Starting from a velocity of about $60$~km\,s$^{-1}$, the limit holds true for approximately the whole range of acceptable $n$ values; this estimate was used for further computations.

The total luminosity of the probable BH lens is estimated according to formula~(\ref{L}) for a mass in the $7.1\pm1.3~M_{\odot}$ interval,
distance $1.58\pm0.18$~kpc, density $n$ from $0.06$ to
$1.3$~cm$^{-3}$, velocity ranging from $40$ to
$100$~km\,s$^{-1}$ and temperature from $200$ to $12\,500$~K. The accretion rate in this case is $\dot m=1.3\times10^{-5}M_{10}n(V^2+c^2_s)^{-3/2}_{16}=1.19{\scriptstyle^{+5.35}_{-0.80}} \times10^{-7}$ in Eddington normalization \cite{2005A&A...440..223B}, which gives a luminosity of $L=5.14{\scriptstyle^{+0.72}_{-0.70}} \times10^{29}$ erg s$^{-1}$, where $V$ is the difference between the BH and interstellar gas velocities. To estimate the latter in the vicinity of the BH, we used the results of \cite{2015ApJ...798...35Z}, who determined the velocities of the medium with respect to the local standard of rest (LSR) from diffuse interstellar bands (DIBs). With a good angular resolution, these velocities in the direction of \mbox{MOA-11-191/OGLE-11-0462} amount to \mbox{$4\pm2$}~km\,s$^{-1}$. This value is lower than the uncertainty level of the velocity of the BH itself, and therefore in this case the gas velocity can be neglected. After transitioning from the total luminosity to $V$-band magnitude in accordance with the accretion spectrum from
\cite{2005A&A...440..223B}, we get an interval of
22.4 to 36.8$^m$. For a
\mbox{$V=60$~km\,s$^{-1}$} velocity of the object, and taking extinction into account, we derive
$m_V=27.87$. The complete list of parameters for the black hole
\mbox{MOA-11-191/OGLE-11-0462} is presented in Table~\ref{tab:parameters}.

\begin{table}[ht!]\begin{center}
\begin{tabular}{ll}\toprule\toprule
  Parameter & Value  \\ 
  \midrule
  
  \scriptsize{Coordinates [$\alpha$,$\delta$ (J2000)]} & \scriptsize{17:51:40.2082, -29:53:26.502} \\
  \scriptsize{Mass [M$_{\odot}$]} & \scriptsize{7.1$\pm$1.3} \\
  \scriptsize{Distance [kpc]} & \scriptsize{1.58$\pm$0.18} \\
  \scriptsize{Transverse velocity [km s$^{-1}$]} & \scriptsize{$\sim$45} \\
  \scriptsize{$^*$Medium density [cm$^{-3}$]} & \scriptsize{0.7$\pm0.4$} \\
  \scriptsize{$^*$Temperature [K]} & \scriptsize{6350$\pm4189$} \\
  \scriptsize{$^*$Sound speed [km s$^{-1}$]} & \scriptsize{6.8$\pm3.6$} \\
  \scriptsize{$^*$Luminosity [erg s$^{-1}$]} & \scriptsize{$5.14{\scriptstyle^{+0.72}_{-0.70}} \times10^{29}$} \\
  \scriptsize{$^*$Accretion rate } & \scriptsize{$1.19{\scriptstyle^{+5.35}_{-0.80}} \times10^{-7}$} \\
  \scriptsize{$^*$Magnitude (for $V \gtrsim60$ km s$^{-1}$)} & \scriptsize{$\gtrsim27.87$} \\
 
  \bottomrule
  
\end{tabular}\caption{\small{Parameters of the black hole MOA-11-191/OGLE-11-0462 and their $\sigma$ level uncertainties. The asterisks mark the quantities obtained in this work}}\label{tab:parameters}
\end{center}\end{table}

In order to estimate the luminosity of the object in other wavelength ranges, as we did above for the optical, we use the results of 
\cite{2005A&A...440..223B}, where spectra were obtained in a wide frequency interval (from radio to gamma) for plasma accreting onto an isolated BH for various properties of both the object and its surrounding medium. Using these computations with the parameters estimated above, we obtain a theoretical spectrum of the probable black hole
\mbox{MOA-11-191/OGLE-11-0462} (presented in Fig.~\ref{fig:detection}), as well as an estimate of the possible flare amplitude (for a given accretion rate its maximum amplitude corresponds to a level of $5.5\%$ of the flux). The $Y$-axis shows the flux density in Jy, the $X$-axis shows the frequencies in Hz. The object is predictably faint in all spectral regions, primarily due to the low accretion rate (about $10^{-7}$), however, we should note that the luminosity of the BH is even lower in other accretion models: for example, it is almost five times lower in \cite{1974Ap&SS..28...45B}, and $60$ times lower in \cite{1982ApJ...255..654I}.

\subsection{The possibility of direct detection of the emission of BH MOA-11-191/OGLE-11-0462}

\begin{figure}[ht!]
   \centering
    \includegraphics[width=1.0\textwidth]{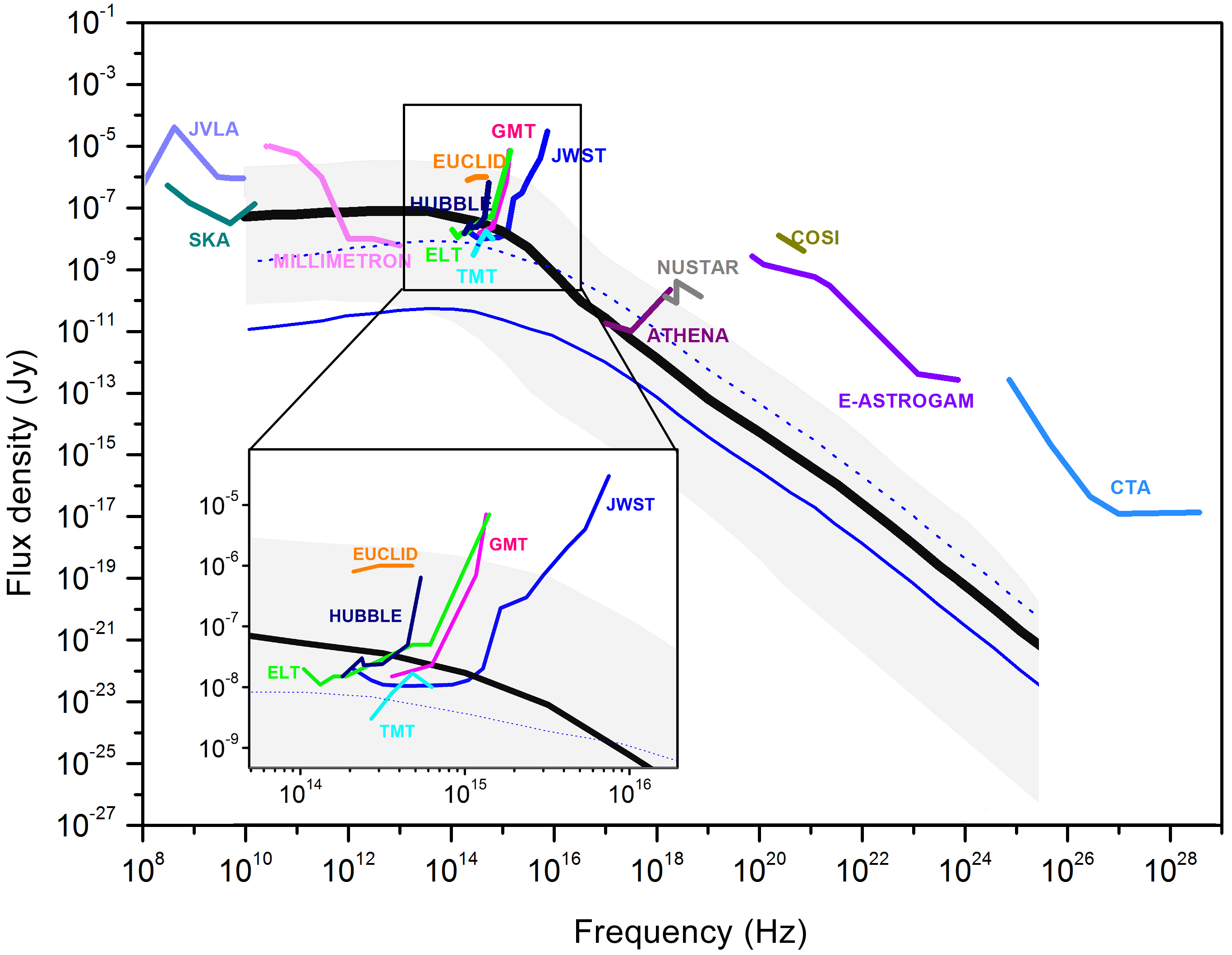}
\caption{\small{Theoretical spectrum of the black hole 
\mbox{MOA-11-191/OGLE-11-0462} with the following parameters:
$M=7.1\pm1.3$~$M_{\odot}$, $D=1.58\pm0.18$~kpc,
$V_{\rm tr}\sim45$~km\,s$^{-1}$,
$n=0.7\pm0.4$~cm$^{-3}$, 
$T=6350\pm4189$~K,
$c_s=6.8\pm3.6$~km\,s$^{-1}$,
\mbox{$L=5.14{\scriptstyle^{+0.72}_{-0.70}}
\times10^{29}$~erg\,s$^{-1}$}, 
\mbox{$\dot
m=1.19{\scriptstyle^{+5.35}_{-0.80}} \times10^{-7}$}. The flux density in Jy is shown as a function of frequency in Hz, the grey area shows the full range of its values corresponding to the acceptable characteristic intervals for the object and interstellar medium. The thick black curve shows the spectrum for $V=60$~km\,s$^{-1}$
(\mbox{$m_V=27.87$}, see text). The maximum amplitude
($5.5\%$) of the possible flares is shown by the thin blue curve, with the dashed curve showing its upper uncertainty. Also shown are the limiting sensitivities of the current and planned observing missions in different frequency ranges. The optical region is emphasized for clarity. }}
\label{fig:detection}
\end{figure}

Detecting emission from the black hole
{MOA-11-191/OGLE-11-0462} is a most important task. Currently, it has not been detected in the optical or any other wavelength range \cite{2022arXiv220113296S}. And although, as was mentioned above, a closer and brighter BH could realistically be detected by modern x-ray observatories, as was discussed by, e.g.,
\cite{2002MNRAS.334..553A,2018MNRAS.477..791T}, as well as in the radio range~--- see
\cite{2019MNRAS.488.2099T,2013MNRAS.430.1538F,2021MNRAS.505.4036S}, the capabilities of modern instruments do not yet allow us to see manifestations of the event horizon of this probable black hole. For instance, the upper sensitivity limit of the x-ray Chandra telescope \cite{2010ApJS..189...37E} is
\mbox{$1.91\times10^{-14}$~erg\,cm$^{-2}$\,s$^{-1}$} in the
\mbox{$1.2\times10^{17}$--$1.7\times10^{18}$~Hz} region, XMM-Newton \cite{2020A&A...641A.136W} can facilitate the detection of emission down to \mbox{$1.52\times10^{-14}$~erg\,cm$^{-2}$\,s$^{-1}$} at
\mbox{$4.8\times10^{16}$--$2.9\times10^{18}$~Hz}, Swift can give about \mbox{$2~\times~10^{-14}$~erg\,cm$^{-2}$~s$^{-1}$}
\cite{2020ApJS..247...54E},  Sperkt~RG~---
\mbox{$2\times10^{-14}$~erg\,cm$^{-2}$\,s$^{-1}$} at 
 \mbox{$(1.2$--$4.8$)$\times10^{17}$}~Hz
\cite{2021A&A...647A...1P}. A comparison of these detection thresholds with the data in Fig.~\ref{fig:detection} clearly shows that direct detection of the hard spectral emission component for this object is currently not possible. At the same time, despite the conclusion of
\cite{2022arXiv220113296S} about no registered light from the lens after the lensing period is over, our estimate shows that the HST could detect it with the  WFC3 camera in the 1--2~micron range (Fig.~3) \cite{2011ApJS..193...27W}.

Studying the regions near the object is an important task. The known kinematics of the lens can be used to estimate the probability of its next microlensing event with another star. Separating the components 
(the distance between them would be about 1 mas) of the background source (star) during the next microlensing event seems possible if one uses an instrument with high spatial resolution, e.g., such as the EHT global interferometer in the millimeter range (20--50~microns)
\cite{2019ApJ...875L...2E,2019ApJ...875L...1E}). The role of such an event in the study of \mbox{MOA-11-191/OGLE-11-0462} is hard to overrate. A rough estimate of its probability amounts to fractions of a percent on a time scale of  10~years and can be refined in a more detailed investigation of the \mbox{MOA-11-191/OGLE-11-0462} vicinity. We believe that such an analysis would also be necessary for other sufficiently long microlensing events.

Let us discuss the possibilities of detecting the emission of matter accreting onto the probable BH - gravitational lens MOA-11-191/OGLE-11-0462 within the framework of contemporary and future missions. Fig.~\ref{fig:detection} compares the sensitivities of the instruments for observations in different frequency regions with the derived spectrum of the probable BH lens. The uncertainties of velocity, distance, mass and interstellar medium density determine the region of acceptable flux values (grey zone in the figure). Such telescopes as SKA, JVLA, Millimetron, JWST, TMT, GMT, ELT, Athena have sufficient limiting sensitivities for detecting the emission of this object in long exposures (the following exposures are shown in the figure: JWST, HST,
ELT, GMT~--- $10^4$~s, ATHENA~--- $10^5$~s, TMT~--- 1~hour, SKA, CTA~--- 100~hours, Millimetron~---
1~day, e-Astrogam~--- 1~year).

We emphasize that the critical test for registering manifestations of an event horizon is the detection of fast variations in the emission of plasma accreting onto a BH \cite{2005A&A...440..223B,
1971SvA....15..377S}. In an approximation of adiabatic heating of the blobs of material ejected from the current sheets, the flare amplitude for a given accretion rate can reach a level of $5.5\%$ of the luminosity in the x-ray range \cite{2005A&A...440..223B}
(thin blue curve in Fig.~\ref{fig:detection}), which gives one a chance of detecting such events. The typical flare durations for adiabatic heating are $\tau_{\nu}\sim r_g/c\sim 10^{-4}-10^{-5}$ seconds
\cite{2005A&A...440..223B}, and a temporal resolution of the order of microseconds combined with a high sensitivity of the Athena telescope will allow the detection of such x-ray flares, the shape of which carries information on the properties of space-time near the event horizon. Note that when magnetic field lines in the current sheets reconnect~--- which is the main mechanism of particle acceleration near the event horizon,~--- their maximal Lorentz factor may reach
$10^{4}$--$10^{5}$ \cite{2005A&A...440..223B}. In these cases, to a distant observer, the intensity of radiation of the electron beams ejected from the current sheets undergoes a multifold increase \cite{1979rpa..book.....R}, and if the orientation of the outbursts is favorable, the flares may be detected by telescopes with nanosecond and microsecond temporal resolutions (CTA, Athena). Thus, there are possibilities of registering the emission of the BH lens in different wavelength ranges and with different temporal characteristics, and therefore, it may be possible to detect observational manifestations of its event horizon. In particular, it should be possible to detect the relatively stable thermal radio emission using the SKA telescopes.

\section{Conclusions}

In this work we analyzed the observed data of a stellar mass black hole, MOA-11-191/OGLE-11-0462, presented by \cite{2022arXiv220113296S}. We estimated the densities, temperatures and sound speeds for the interstellar medium in its vicinity. Using the observed magnitude limitations of the BH, its probable total velocity was estimated. The derived values were used to compute the accretion rate and total luminosity of the object. Its spectrum was constructed in various frequency domains, and the possible flare levels were estimated. Currently, no emission from MOA-11-191/OGLE-11-0462 has been detected in the optical or any other wavelength regions
\cite{2022arXiv220113296S}. However, our estimate shows that the HST WFC3 camera should be able to register it
\cite{2011ApJS..193...27W}. Telescopes such as SKA, JVLA,
Millimetron, JWST, TMT, GMT, ELT,
Athena are sufficiently sensitive for the detection of the object's radiation at long exposures. Its thermal radio emission may be detected with the SKA telescopes. The critical test for registering the manifestations of an event horizon is the detection of fast variations in the emission of plasma accreting onto the BH \cite{2005A&A...440..223B,
1971SvA....15..377S}. The amplitude of the flares may reach a level of 
$5.5\%$ of the x-ray luminosity
\cite{2005A&A...440..223B}, which makes it possible to detect such events with an instrument of microsecond temporal resolution and high sensitivity (Athena). Telescopes with nanosecond and microsecond temporal resolution (CTA, Athena) can detect flares in the case of a fortunate orientation of the outbursts (when the radiation intensity of the ejected electron beams increases for an observer \cite{1979rpa..book.....R}). Possibilities thus exist for registering the emission of the probable BH lens
MOA-11-191/OGLE-11-0462 at different wavelengths and for different temporal characteristics, and therefore, the manifestations of its event horizon can be detected.\\

\noindent\textbf{Acknowledgments}\\
This work was performed within the grant number 075-15-2022-262 of the Ministry of science and higher education of the Russian Federation (13.MNPMU.21.0003).

\bibliographystyle{elsarticle-num}
\bibliography{references.bib}

\begin{thebibliography}{10}
\expandafter\ifx\csname url\endcsname\relax
  \def\url#1{\texttt{#1}}\fi
\expandafter\ifx\csname urlprefix\endcsname\relax\def\urlprefix{URL }\fi
\expandafter\ifx\csname href\endcsname\relax
  \def\href#1#2{#2} \def\path#1{#1}\fi

\bibitem{2019ApJ...885....1W}
G.~{Wiktorowicz}, {\L}.~{Wyrzykowski}, M.~{Chruslinska}, J.~{Klencki}, K.~A.
  {Rybicki}, K.~{Belczynski}, {Populations of Stellar-mass Black Holes from
  Binary Systems}, \apj 885~(1) (2019) 1.
\newblock \href {http://arxiv.org/abs/1907.11431} {\path{arXiv:1907.11431}},
  \href {https://doi.org/10.3847/1538-4357/ab45e6}
  {\path{doi:10.3847/1538-4357/ab45e6}}.

\bibitem{2008AdSpR..42..523B}
G.~{Beskin}, A.~{Biryukov}, S.~{Karpov}, V.~{Plokhotnichenko}, V.~{Debur},
  {Observational appearances of isolated stellar-mass black hole accretion
  Theory and observations}, Advances in Space Research 42~(3) (2008) 523--532.
\newblock \href {http://arxiv.org/abs/0709.2552} {\path{arXiv:0709.2552}},
  \href {https://doi.org/10.1016/j.asr.2007.03.104}
  {\path{doi:10.1016/j.asr.2007.03.104}}.

\bibitem{1944MNRAS.104..273B}
H.~{Bondi}, F.~{Hoyle}, {On the mechanism of accretion by stars}, \mnras 104
  (1944) 273.
\newblock \href {https://doi.org/10.1093/mnras/104.5.273}
  {\path{doi:10.1093/mnras/104.5.273}}.

\bibitem{2005A&A...440..223B}
G.~M. {Beskin}, S.~V. {Karpov}, {Low-rate accretion onto isolated stellar-mass
  black holes}, \aap 440~(1) (2005) 223--238.
\newblock \href {http://arxiv.org/abs/astro-ph/0403649}
  {\path{arXiv:astro-ph/0403649}}, \href
  {https://doi.org/10.1051/0004-6361:20040572}
  {\path{doi:10.1051/0004-6361:20040572}}.

\bibitem{1992ans..book.....L}
V.~M. {Lipunov}, {Astrophysics of Neutron Stars}, 1992.

\bibitem{1981cfgr.book.....K}
W.~J. {Kaufmann}, {The cosmic frontiers of general relativity.}, 1981.

\bibitem{1986UsFiN.148..393D}
I.~G. {Dymnikova}, {Motion of particles and photons in the gravitational field
  of a rotating body.}, Uspekhi Fizicheskikh Nauk 148~(3) (1986) 393--432.

\bibitem{1971SvA....15..377S}
V.~F. {Shvartsman}, {Halos around ``Black Holes''.}, \sovast 15 (1971) 377.

\bibitem{1974Ap&SS..28...45B}
G.~S. {Bisnovatyi-Kogan}, A.~A. {Ruzmaikin}, {The Accretion of Matter by a
  Collapsing Star in the Presence of a Magnetic Field}, \apss 28~(1) (1974)
  45--59.
\newblock \href {https://doi.org/10.1007/BF00642237}
  {\path{doi:10.1007/BF00642237}}.

\bibitem{1975A&A....44...59M}
P.~{Meszaros}, {Radiation from spherical accretion onto black holes.}, \aap
  44~(1) (1975) 59--68.

\bibitem{1982ApJ...255..654I}
J.~R. {Ipser}, R.~H. {Price}, {Synchrotron radiation from spherically accreting
  black holes}, \apj 255 (1982) 654--673.
\newblock \href {https://doi.org/10.1086/159866} {\path{doi:10.1086/159866}}.

\bibitem{1977SoSAO..19....5S}
V.~F. {Shvartsman}, {The MANIA [Multichannel Analysis of Nanosecond Intensity
  Alterations] experiment. Astrophysical problems, mathematical methods,
  instrumentation complex, results of the first observations.}, Soobshcheniya
  Spetsial'noj Astrofizicheskoj Observatorii 19 (1977) 5--38.

\bibitem{1989Afz....31..457S}
V.~F. {Shvartsman}, G.~M. {Beskin}, S.~A. {Pustil'Nik}, {The results of search
  for superrapid optical variability of radio objects with continuous optical
  spectra.}, Astrofizika 31 (1989) 457--465.

\bibitem{1989SvAL...15..145S}
V.~F. {Shvartsman}, G.~M. {Beskin}, S.~N. {Mitronova}, {A Search for
  0.5-MICROSECOND to 40-SECOND Optical Variability in DC White Dwarfs}, Soviet
  Astronomy Letters 15 (1989) 145.

\bibitem{2010AJ....139..390P}
R.~M. {Plotkin}, S.~F. {Anderson}, W.~N. {Brandt}, A.~M. {Diamond-Stanic},
  X.~{Fan}, P.~B. {Hall}, A.~E. {Kimball}, M.~W. {Richmond}, {et al.},
  {Optically Selected BL Lacertae Candidates from the Sloan Digital Sky Survey
  Data Release Seven}, \aj 139~(2) (2010) 390--414.
\newblock \href {http://arxiv.org/abs/0911.0423} {\path{arXiv:0911.0423}},
  \href {https://doi.org/10.1088/0004-6256/139/2/390}
  {\path{doi:10.1088/0004-6256/139/2/390}}.

\bibitem{2010AstL...36..116C}
E.~G. {Chmyreva}, G.~M. {Beskin}, A.~V. {Biryukov}, {Search for pairs of
  isolated radio pulsars{\textemdash}Components in disrupted binary systems},
  Astronomy Letters 36~(2) (2010) 116--133.
\newblock \href {http://arxiv.org/abs/1203.2836} {\path{arXiv:1203.2836}},
  \href {https://doi.org/10.1134/S1063773710020040}
  {\path{doi:10.1134/S1063773710020040}}.

\bibitem{2022AstBu..77...65C}
L.~{Chmyreva}, G.~M. {Beskin}, {Peculiar Objects in the Birthplaces of Radio
  Pulsars{\textemdash}Stellar-Mass Black Hole Candidates}, Astrophysical
  Bulletin 77~(1) (2022) 65--77.
\newblock \href {https://doi.org/10.1134/S1990341322010035}
  {\path{doi:10.1134/S1990341322010035}}.

\bibitem{2002MNRAS.334..553A}
E.~{Agol}, M.~{Kamionkowski}, {X-rays from isolated black holes in the Milky
  Way}, \mnras 334~(3) (2002) 553--562.
\newblock \href {http://arxiv.org/abs/astro-ph/0109539}
  {\path{arXiv:astro-ph/0109539}}, \href
  {https://doi.org/10.1046/j.1365-8711.2002.05523.x}
  {\path{doi:10.1046/j.1365-8711.2002.05523.x}}.

\bibitem{2018MNRAS.477..791T}
D.~{Tsuna}, N.~{Kawanaka}, T.~{Totani}, {X-ray detectability of accreting
  isolated black holes in our Galaxy}, \mnras 477~(1) (2018) 791--801.
\newblock \href {http://arxiv.org/abs/1801.04667} {\path{arXiv:1801.04667}},
  \href {https://doi.org/10.1093/mnras/sty699}
  {\path{doi:10.1093/mnras/sty699}}.

\bibitem{2012MNRAS.427..589B}
M.~V. {Barkov}, D.~V. {Khangulyan}, S.~B. {Popov}, {Jets and gamma-ray emission
  from isolated accreting black holes}, \mnras 427~(1) (2012) 589--594.
\newblock \href {http://arxiv.org/abs/1209.0293} {\path{arXiv:1209.0293}},
  \href {https://doi.org/10.1111/j.1365-2966.2012.22029.x}
  {\path{doi:10.1111/j.1365-2966.2012.22029.x}}.

\bibitem{2005MNRAS.360L..30M}
T.~J. {Maccarone}, {Using radio emission to detect isolated and quiescent
  accreting black holes}, \mnras 360~(1) (2005) L30--L34.
\newblock \href {http://arxiv.org/abs/astro-ph/0503097}
  {\path{arXiv:astro-ph/0503097}}, \href
  {https://doi.org/10.1111/j.1745-3933.2005.00039.x}
  {\path{doi:10.1111/j.1745-3933.2005.00039.x}}.

\bibitem{2016PhRvX...6d1015A}
B.~P. {Abbott}, R.~{Abbott}, T.~D. {Abbott}, M.~R. {Abernathy}, F.~{Acernese},
  {et al.}, {Binary Black Hole Mergers in the First Advanced LIGO Observing
  Run}, Physical Review X 6~(4) (2016) 041015.
\newblock \href {http://arxiv.org/abs/1606.04856} {\path{arXiv:1606.04856}},
  \href {https://doi.org/10.1103/PhysRevX.6.041015}
  {\path{doi:10.1103/PhysRevX.6.041015}}.

\bibitem{2016ApJ...818L..22A}
B.~P. {Abbott}, R.~{Abbott}, T.~D. {Abbott}, M.~R. {Abernathy}, F.~{Acernese},
  K.~{Ackley}, C.~{Adams}, T.~{Adams}, P.~{Addesso}, {et al.}, {Astrophysical
  Implications of the Binary Black-hole Merger GW150914}, \apjl 818~(2) (2016)
  L22.
\newblock \href {http://arxiv.org/abs/1602.03846} {\path{arXiv:1602.03846}},
  \href {https://doi.org/10.3847/2041-8205/818/2/L22}
  {\path{doi:10.3847/2041-8205/818/2/L22}}.

\bibitem{2019MNRAS.488.2099T}
D.~{Tsuna}, N.~{Kawanaka}, {Radio emission from accreting isolated black holes
  in our galaxy}, \mnras 488~(2) (2019) 2099--2107.
\newblock \href {http://arxiv.org/abs/1907.00792} {\path{arXiv:1907.00792}},
  \href {https://doi.org/10.1093/mnras/stz1809}
  {\path{doi:10.1093/mnras/stz1809}}.

\bibitem{2013MNRAS.430.1538F}
R.~P. {Fender}, T.~J. {Maccarone}, I.~{Heywood}, {The closest black holes},
  \mnras 430~(3) (2013) 1538--1547.
\newblock \href {http://arxiv.org/abs/1301.1341} {\path{arXiv:1301.1341}},
  \href {https://doi.org/10.1093/mnras/sts688}
  {\path{doi:10.1093/mnras/sts688}}.

\bibitem{2021MNRAS.505.4036S}
F.~{Scarcella}, D.~{Gaggero}, R.~{Connors}, J.~{Manshanden}, M.~{Ricotti},
  G.~{Bertone}, {Multiwavelength detectability of isolated black holes in the
  Milky Way}, \mnras 505~(3) (2021) 4036--4047.
\newblock \href {http://arxiv.org/abs/2012.10421} {\path{arXiv:2012.10421}},
  \href {https://doi.org/10.1093/mnras/stab1533}
  {\path{doi:10.1093/mnras/stab1533}}.

\bibitem{2019MNRAS.489.2038I}
P.~B. {Ivanov}, V.~N. {Lukash}, S.~V. {Pilipenko}, M.~S. {Pshirkov}, {Search
  for isolated Galactic Centre stellar mass black holes in the IR and sub-mm
  range}, \mnras 489~(2) (2019) 2038--2048.
\newblock \href {http://arxiv.org/abs/1905.04923} {\path{arXiv:1905.04923}},
  \href {https://doi.org/10.1093/mnras/stz2206}
  {\path{doi:10.1093/mnras/stz2206}}.

\bibitem{2016ApJ...830...41L}
J.~R. {Lu}, E.~{Sinukoff}, E.~O. {Ofek}, A.~{Udalski}, S.~{Kozlowski}, {A
  Search For Stellar-mass Black Holes Via Astrometric Microlensing}, \apj
  830~(1) (2016) 41.
\newblock \href {http://arxiv.org/abs/1607.08284} {\path{arXiv:1607.08284}},
  \href {https://doi.org/10.3847/0004-637X/830/1/41}
  {\path{doi:10.3847/0004-637X/830/1/41}}.

\bibitem{2022arXiv220201903L}
C.~Y. {Lam}, J.~R. {Lu}, A.~{Udalski}, I.~{Bond}, D.~P. {Bennett},
  J.~{Skowron}, P.~{Mroz}, R.~{Poleski}, {et al}, {An isolated mass gap black
  hole or neutron star detected with astrometric microlensing}, arXiv e-prints
  (2022) arXiv:2202.01903\href {http://arxiv.org/abs/2202.01903}
  {\path{arXiv:2202.01903}}.

\bibitem{2022arXiv220113296S}
K.~C. {Sahu}, J.~{Anderson}, S.~{Casertano}, H.~E. {Bond}, A.~{Udalski}, {et
  al}, {An Isolated Stellar-Mass Black Hole Detected Through Astrometric
  Microlensing}, arXiv e-prints (2022) arXiv:2201.13296\href
  {http://arxiv.org/abs/2201.13296} {\path{arXiv:2201.13296}}.

\bibitem{1952MNRAS.112..195B}
H.~{Bondi}, {On spherically symmetrical accretion}, \mnras 112 (1952) 195.
\newblock \href {https://doi.org/10.1093/mnras/112.2.195}
  {\path{doi:10.1093/mnras/112.2.195}}.

\bibitem{2019ApJ...887...93G}
G.~M. {Green}, E.~{Schlafly}, C.~{Zucker}, J.~S. {Speagle}, D.~{Finkbeiner}, {A
  3D Dust Map Based on Gaia, Pan-STARRS 1, and 2MASS}, \apj 887~(1) (2019) 93.
\newblock \href {http://arxiv.org/abs/1905.02734} {\path{arXiv:1905.02734}},
  \href {https://doi.org/10.3847/1538-4357/ab5362}
  {\path{doi:10.3847/1538-4357/ab5362}}.

\bibitem{2009MNRAS.400.2050G}
T.~{G{\"u}ver}, F.~{{\"O}zel}, {The relation between optical extinction and
  hydrogen column density in the Galaxy}, \mnras 400~(4) (2009) 2050--2053.
\newblock \href {http://arxiv.org/abs/0903.2057} {\path{arXiv:0903.2057}},
  \href {https://doi.org/10.1111/j.1365-2966.2009.15598.x}
  {\path{doi:10.1111/j.1365-2966.2009.15598.x}}.

\bibitem{1981sss..book..265B}
N.~G. {Bochkarev}, {The interstellar medium and star formation}, 1981, pp.
  265--325.

\bibitem{2022wfci.book...14D}
L.~{Dressel}, {WFC3 Instrument Handbook for Cycle 30 v. 14}, in: WFC3
  Instrument Handbook for Cycle 30 v. 14, Vol.~14, 2022, p.~14.

\bibitem{2015ApJ...798...35Z}
G.~{Zasowski}, B.~{M{\'e}nard}, D.~{Bizyaev}, D.~A.
  {Garc{\'\i}a-Hern{\'a}ndez}, A.~E. {Garc{\'\i}a P{\'e}rez}, M.~R. {Hayden},
  J.~{Holtzman}, J.~A. {Johnson}, K.~{Kinemuchi}, S.~R. {Majewski}, D.~L.
  {Nidever}, M.~{Shetrone}, J.~C. {Wilson}, {Mapping the Interstellar Medium
  with Near-infrared Diffuse Interstellar Bands}, \apj 798~(1) (2015) 35.
\newblock \href {http://arxiv.org/abs/1406.1195} {\path{arXiv:1406.1195}},
  \href {https://doi.org/10.1088/0004-637X/798/1/35}
  {\path{doi:10.1088/0004-637X/798/1/35}}.

\bibitem{2010ApJS..189...37E}
I.~N. {Evans}, F.~A. {Primini}, K.~J. {Glotfelty}, C.~S. {Anderson}, {et al.},
  {The Chandra Source Catalog}, \apjs 189~(1) (2010) 37--82.
\newblock \href {http://arxiv.org/abs/1005.4665} {\path{arXiv:1005.4665}},
  \href {https://doi.org/10.1088/0067-0049/189/1/37}
  {\path{doi:10.1088/0067-0049/189/1/37}}.

\bibitem{2020A&A...641A.136W}
N.~A. {Webb}, M.~{Coriat}, I.~{Traulsen}, J.~{Ballet}, C.~{Motch}, F.~J.
  {Carrera}, F.~{Koliopanos}, J.~{Authier}, I.~{de la Calle}, {et al.}, {The
  XMM-Newton serendipitous survey. IX. The fourth XMM-Newton serendipitous
  source catalogue}, \aap 641 (2020) A136.
\newblock \href {http://arxiv.org/abs/2007.02899} {\path{arXiv:2007.02899}},
  \href {https://doi.org/10.1051/0004-6361/201937353}
  {\path{doi:10.1051/0004-6361/201937353}}.

\bibitem{2020ApJS..247...54E}
P.~A. {Evans}, K.~L. {Page}, J.~P. {Osborne}, A.~P. {Beardmore},
  R.~{Willingale}, D.~N. {Burrows}, J.~A. {Kennea}, M.~{Perri}, M.~{Capalbi},
  G.~{Tagliaferri}, S.~B. {Cenko}, {2SXPS: An Improved and Expanded Swift X-Ray
  Telescope Point-source Catalog}, \apjs 247~(2) (2020) 54.
\newblock \href {http://arxiv.org/abs/1911.11710} {\path{arXiv:1911.11710}},
  \href {https://doi.org/10.3847/1538-4365/ab7db9}
  {\path{doi:10.3847/1538-4365/ab7db9}}.

\bibitem{2021A&A...647A...1P}
P.~{Predehl}, R.~{Andritschke}, V.~{Arefiev}, V.~{Babyshkin}, O.~{Batanov},
  W.~{Becker}, H.~{B{\"o}hringer}, A.~{Bogomolov}, {et al.}, {The eROSITA X-ray
  telescope on SRG}, \aap 647 (2021) A1.
\newblock \href {http://arxiv.org/abs/2010.03477} {\path{arXiv:2010.03477}},
  \href {https://doi.org/10.1051/0004-6361/202039313}
  {\path{doi:10.1051/0004-6361/202039313}}.

\bibitem{2011ApJS..193...27W}
R.~A. {Windhorst}, S.~H. {Cohen}, N.~P. {Hathi}, P.~J. {McCarthy}, J.~{Ryan},
  R.~E., H.~{Yan}, I.~K. {Baldry}, {et al.}, {The Hubble Space Telescope Wide
  Field Camera 3 Early Release Science Data: Panchromatic Faint Object Counts
  for 0.2-2 {\ensuremath{\mu}}m Wavelength}, \apjs 193~(2) (2011) 27.
\newblock \href {http://arxiv.org/abs/1005.2776} {\path{arXiv:1005.2776}},
  \href {https://doi.org/10.1088/0067-0049/193/2/27}
  {\path{doi:10.1088/0067-0049/193/2/27}}.

\bibitem{2019ApJ...875L...2E}
{Event Horizon Telescope Collaboration}, K.~{Akiyama}, A.~{Alberdi}, W.~{Alef},
  K.~{Asada}, R.~{Azulay}, {et al.}, {First M87 Event Horizon Telescope
  Results. II. Array and Instrumentation}, \apjl 875~(1) (2019) L2.
\newblock \href {http://arxiv.org/abs/1906.11239} {\path{arXiv:1906.11239}},
  \href {https://doi.org/10.3847/2041-8213/ab0c96}
  {\path{doi:10.3847/2041-8213/ab0c96}}.

\bibitem{2019ApJ...875L...1E}
{Event Horizon Telescope Collaboration}, {First M87 Event Horizon Telescope
  Results. I. The Shadow of the Supermassive Black Hole}, \apjl 875~(1) (2019)
  L1.
\newblock \href {http://arxiv.org/abs/1906.11238} {\path{arXiv:1906.11238}},
  \href {https://doi.org/10.3847/2041-8213/ab0ec7}
  {\path{doi:10.3847/2041-8213/ab0ec7}}.

\bibitem{1979rpa..book.....R}
G.~B. {Rybicki}, A.~P. {Lightman}, {Radiative processes in astrophysics}, 1979.

\end{thebibliography}
 
\end{document}